\documentclass[twocolumn,notitlepage,prl,superscriptaddress,showpacs]{revtex4-2}
\usepackage{hyperref}
\usepackage{mathtools}
\usepackage[usenames,dvipsnames]{color}
\usepackage{amsmath}
\usepackage[english]{babel}
\usepackage{amssymb}
\usepackage{graphicx}
\usepackage{epstopdf}
\usepackage{xcolor}
\usepackage{tabularx}
\usepackage{subfig}
\usepackage{placeins}
\usepackage{afterpage}
\usepackage{dblfloatfix} 
\usepackage{float}

\begin{document}

\title{Evaluating second-order phase transitions with Diagrammatic Monte Carlo: \\
N\'{e}el Transition in the doped three-dimensional Hubbard model}

\author{Connor Lenihan}
\email{connor.lenihan@kcl.ac.uk}
\affiliation{Department of Physics, King's College London, Strand, London WC2R 2LS, UK}

\author{Aaram J. Kim}
\affiliation{Department of Physics, King's College London, Strand, London WC2R 2LS, UK}
\affiliation{Department of Physics, University of Fribourg, Chemin du Mus\'ee 3, 1700 Fribourg, Switzerland}

\author{Fedor \v{S}imkovic IV.}
\affiliation{Department of Physics, King's College London, Strand, London WC2R 2LS, UK}
\affiliation{CPHT, CNRS, \'Ecole Polytechnique, Institut Polytechnique de Paris, Route de Saclay, 91128 Palaiseau, France} 
\affiliation{Coll\`{e}ge de France, 11 place Marcelin Berthelot, 75005 Paris, France}

\author{Evgeny Kozik}
\email{evgeny.kozik@kcl.ac.uk}
\affiliation{Department of Physics, King's College London, Strand, London WC2R 2LS, UK}

\begin{abstract}
Diagrammatic Monte Carlo---the technique for numerically exact summation of all Feynman diagrams to high orders---offers a unique unbiased probe of continuous phase transitions. Being formulated directly in the thermodynamic limit, the diagrammatic series is bound to diverge and is not resummable at the transition due to the non-analyticity of physical observables. This enables the detection of the transition with controlled error bars from an analysis of the series coefficients alone, avoiding the challenge of evaluating physical observables near the transition. We demonstrate this technique by the example of the N\'eel transition in the $3d$ Hubbard model. At half-filling and higher temperatures, the method matches the accuracy of state-of-the-art finite-size techniques, but surpasses it at low temperatures and allows us to map the phase diagram in the doped regime, where finite-size techniques struggle from the fermion sign problem. At low temperatures and sufficient doping, the transition to an incommensurate spin density wave state is observed.
\end{abstract}


\maketitle

The universal technique of Feynman diagrams~\cite{AGD} has become a precise computational tool for quantum many-body systems in and out of equilibrium. A wealth of diagrammatic Monte Carlo (DiagMC)~\cite{Prokofev1998, VanHoucke2010,Kozik2010} techniques, developed recently for diverse physical systems and types of Feynman diagrammatic expansions~\cite{VanHoucke2012, deng2015emergent, Gukelberge2014nematic, profumo2015, Iskakov2016df, Gukelberger2017df, wu2017, Rossi2016det, moutenet2018determinant, Simkovic2019, Chen2019, Bertrand2019, Kim2020spin_charge, Vandelli2020db, simkovicefficient2020, rossi2020renormalized, li2020diagrammatic, vsimkovic2021weak} enables numerically exact summation of the series to very high orders, allowing one to compute physical observables in strongly correlated regimes with controlled accuracy. The key advantage of the DiagMC approach is that Feynman diagrams can be formulated immediately in the thermodynamic limit (TDL), the cutoff of the expansion order being the only parameter controlling the accuracy. This has enabled, in particular, unbiased solutions to problems where the correlations are intrinsically long-ranged, such as that of the metal-to-insulator crossover in the $2d$ Hubbard model \cite{Fedor2020,Kim2020spin_charge, Connor2021entropy, vsimkovic2021weak}. 

There is, however, an additional consequence, a feature unavailable in finite-size techniques, that has so far not been used or explored. At the points of continuous (second-order) phase transitions, defined only in the TDL, thermodynamic potentials and relevant susceptibilities are non-analytic functions of their variables. The non-analyticity, in turn, implies that perturbative---in the powers of the coupling strength---expansions for these functions constructed in the TDL must diverge beyond the phase transition. Moreover, the character of this divergence is controlled by the universality class of the transition.

Here we demonstrate that second-order phase transitions can be reliably detected and characterised by the high-order asymptotic behavior of the diagrammatic series expansion evaluated by DiagMC. This approach to criticality does not require reconstructing physical observables from the series, while approaching the problem from the TDL leads to a technique which is fundamentally different from and complementary to the finite-size scaling approach of quantum Monte Carlo calculations~\cite{Staudt2000,Kent2005,Kozik2013Neel}. We illustrate our approach by its application to the problem of the antiferromagnetic (AFM, N\'eel) transition in the repulsive $3d$ Hubbard model, which is important in its own right.

The Hubbard model~\cite{hubbard1963electron} is the simplest model for correlated fermions, the main benchmark for controlled numerical methods~\cite{leblanc2015solutions}, and the paradigm for fundamental many-body mechanisms, including unconventional superconductivity and quantum magnetism~\cite{anderson1963theory, anderson1997theory}. Its Hamiltonian is given by:
\begin{align}
	H = -  t\sum_{\left<i,j\right>, \sigma}  \hat{c}^{\dagger}_{i, \sigma} \hat{c}^{}_{j, \sigma}
+ U \sum_{i} \hat{n}_{i, \uparrow} \hat{n}_{i \downarrow} -\mu \sum_{i, \sigma} \hat{n}_{i, \sigma},
    \label{Hubbard}
\end{align}
where $\mu$ is the chemical potential, $U$ the interaction strength, $t$ the hopping, 
$\hat{c}^{\dagger}_{i, \sigma}$ ($\hat{c}_{i, \sigma}$)
create (annihilate) a fermion with the spin $\sigma$
on the site $i$ and $\hat{n}_{i, \sigma}=\hat{c}^{\dagger}_{i, \sigma} \hat{c}_{i, \sigma}$. It is used as a conceptual model of high-$T_c$ superconducting cuprates, where the layered structure makes the $2d$ physics dominant, but the weak coupling between the layers stabilises long-range order. In particular, the AFM state is part of the cuprates' phase diagram in a range of doping near half-filling (the average number of electrons per lattice site $n=1$). A number of perovskite oxide compounds, such as titanates, vanadates and nickelates are also well described by the $3d$ Hubbard model~\cite{klebel2021anisotropy}. The $3d$ Hubbard model and its AFM physics in particular, alongside its $2d$ counterpart, is therefore a crucial baseline for understanding more realistic systems~\cite{RevModPhys.70.1039,klebel2021anisotropy}. Much of additional recent interest in the Hubbard model has been spurred by the ability to simulate the model with ultracold atoms in optical lattices (UCA)~\cite{TARRUELL2018365, Bloch:2005uv, Lewenstein:2007hr, Bloch2012, Greif:2015bg, Hart2015, Cheuk:2016kq, Mazurenko2017, Brown:2017dy, Nichols:2019iq} and the demonstration of the AFM state in a UCA system ~\cite{Mazurenko2017}, which provides an interface between theory and experiment and allows the results of theoretical predictions to be validated and the exploration of regimes which are  challenging for theory.

However, predicting the phase diagram of the model has proven to be a challenge. In $2d$, the development of the (quasi-)AFM state, responsible for the metal-to-insulator crossover~\cite{Lichtenstein2000afm_sc, Schafer:2015jg, Rohringer:2016jt}, has only recently come within reach of controlled methods~\cite{Fedor2020, Kim2020spin_charge,  Connor2021entropy, vsimkovic2021weak}. This regime has become a test-bed for novel numerical approaches \cite{leblanc2015solutions,schaefertracking2021}, while its $3d$ half-filled counterpart has served for benchmarking approximate methods for real materials \cite{iskakov2021phase}.  In $3d$, unbiased simulations have been facilitated by the power-law scaling of the correlation length near the N\'eel transition---in contrast to the exponential scaling in the $T \to 0$ limit in $2d$---and the N\'eel temperature $T_N$ at half-filling has been mapped out with controlled error bars~\cite{Staudt2000,Kent2005,Kozik2013Neel}. Specifically, due to the absence of the fermion sign problem~\cite{loh1990sign} at half-filling, simulations of sufficiently large systems are feasible for finite-size methods, such as quantum Monte Carlo (QMC)~\cite{Staudt2000}, dynamical cluster approximation (DCA)~\cite{Kent2005},
%
%
and determinant diagrammatic Monte Carlo (DDMC)~\cite{Kozik2013Neel}. The critical point can be determined in a controlled way as soon as the universal critical scaling of relevant observables with the system size has been reached (the so-called finite-size scaling). In the doped regime, however, the sign problem leads to an additional exponential scaling of the computational time with the system size and inverse temperature~\cite{troyer2005sign}, precluding access to large enough systems for a reliable extrapolation to the TDL. Thus, little is known conclusively about the doped $3d$ Hubbard model, while state-of-the-art field-theoretic calculations by the D$\Gamma$A method~\cite{Schaefer2017} predict a transition between an AFM and an incommensurate spin density wave (SDW) state at high enough doping.

\begin{figure}[h!]
\includegraphics[width=1.0\columnwidth]{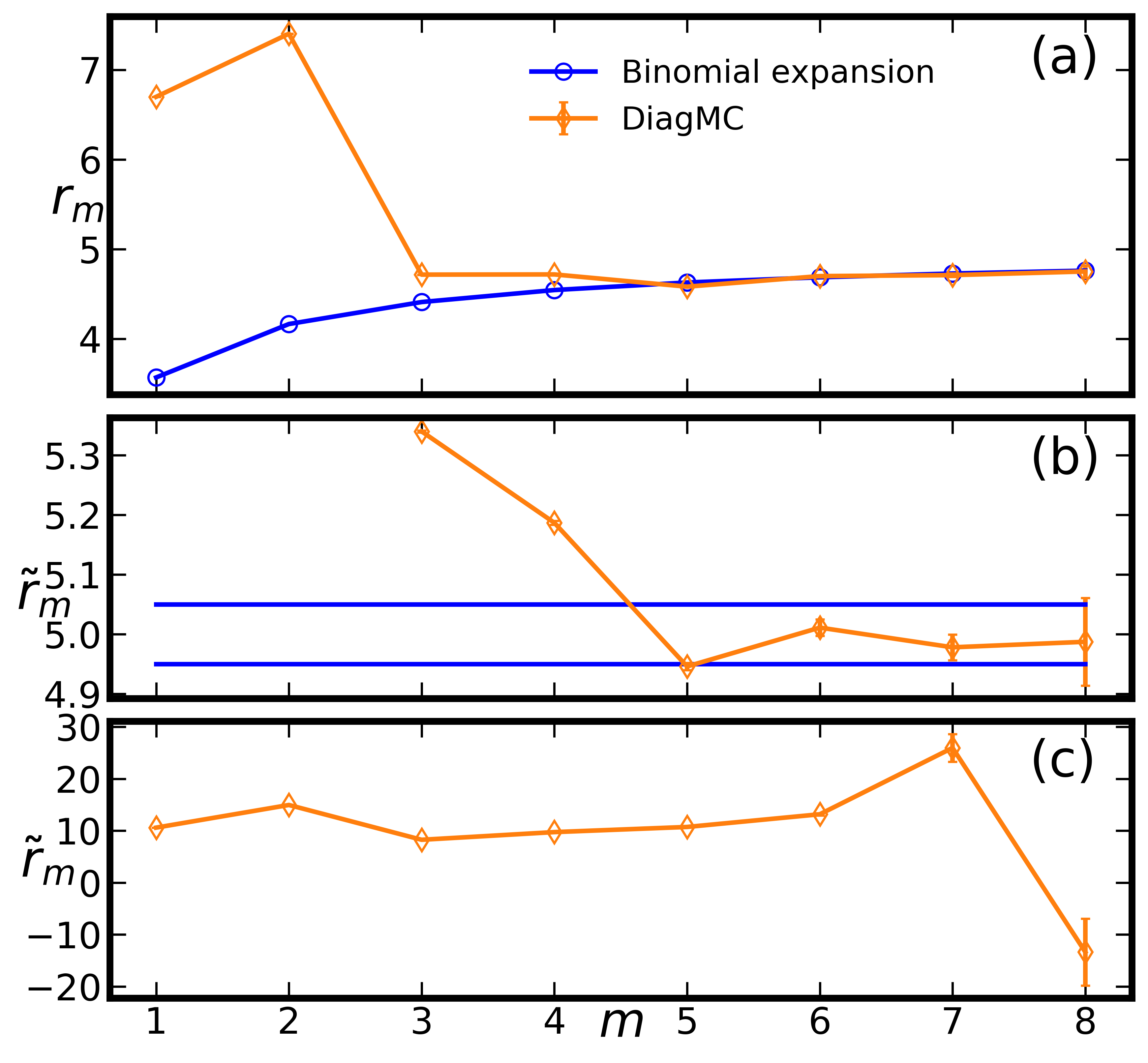}
\caption{Demonstration of our method for locating the critical point $U_c$ based on Eq.~\ref{U_c}: (a) the ratio of consecutive series coefficients $r_m$ vs. order $m$ for $S(\boldsymbol{\pi})$ of the $3d$ Hubbard model at $T=0.25t$, $n=1.0$ (orange) approaches the asymptote $\frac{m U_c}{m - 1 + \gamma}$ with $U_c=5.0t$ (blue), expected from the critical scaling $S(\boldsymbol{\pi}) \propto (U_c - U)^{-\gamma}$; (b) [same data as (a)] in practice, the adjusted ratio $\tilde{r}_m=\frac{m - 1 + \gamma}{m} r_m$ is used to obtain $U_c$ from the location of the plateau at large orders, the blue lines are the upper and lower error bounds we put on $U_c$; (c) the adjusted ratio for a temperature ($T=0.5t$) where there is no phase transition at any $U$ and hence no plateau.}
\label{adjusted_ratio}
\end{figure}

In this paper we use the new technique to map out the AFM phase transition in the $3d$ Hubbard model at half-filling, where we validate it by benchmarking against the established unbiased data, and equally in the doped regime, for which no controlled results currently exist. We find for $T=0.05t$ that beyond $\sim 10\%$ doping the N\'eel transition is to an incommensurate SDW state, in qualitative agreement with the D$\Gamma$A findings~\cite{Schaefer2017}. Our controlled results for the spin-spin correlation function illustrate the nature of underpinning correlations and are amenable to experimental validation with UCA in optical lattices. 


For an observable $\mathcal{A}$, DiagMC stochastically sums---without any approximations---all Feynman diagrams comprising the coefficients $a_m$ of the perturbative expansion $\mathcal{A}=\sum_m a_m U^m$
up to some truncation order $m_*$, at which the Monte Carlo error bars become impractically large. Due to its formulation in the TDL, the DiagMC approach circumvents the fundamental complexity of the negative sign problem. Indeed, it was demonstrated~\cite{Rossi2017ccp, Kim2021homotopy} that the calculation time generically scales only polynomially with the inverse of the desired error bound, provided the series converges. The series can sometimes diverge due to singularities in the complex plane of $U$---which limit the convergence radius and are quite typical~\cite{profumo2015, Rossi2016det, Simkovic2019, Fedor2020, Kim2020spin_charge}---but as long as they are not on the real axis with $\operatorname{Re}\! U > 0$ the series convergence properties can be improved by altering the starting points of the perturbative expansions \cite{profumo2015, wu2017, rossi2020renormalized, simkovicefficient2020, spada2021highorder, vsimkovic2021weak}, or more generally, a homotopy of the effective action~\cite{Kim2021homotopy}. However, at a point $U_c$ of a continuous phase transition, the exact function $\mathcal{A}(U)$ (with other variables fixed) exhibits a physical non-analyticity with a power-law singularity, $\mathcal{A}(U) \propto (U_c - U)^{- \gamma}$ ($\gamma$ is real) for $U \to U_c$, and the series must diverge at least for $U > U_c$. This fundamental failure of the DiagMC approach generally allows one to detect the transition even when its nature is unknown \textit{a priori}. When the symmetry broken-state is known, constructing the expansion about its mean-field solution has been shown to enable DiagMC calculations across the transition~\cite{spada2021highorder}; here we argue that the breakdown of the technique is a useful computational tool in itself.

The location of the singularity $U_c=U_c(T)$ can be identified from the ratios of adjacent coefficients, $U_c = \lim_{n\to\infty}{{\frac{a_{n-1}}{a_{n}}}}$, for a fixed temperature $T$. When the exponent $\gamma$ is known, the asymptotic form of the binomial expansion can be used to improve the accuracy of locating the singularity, 
\begin{equation}
U_c = \lim_{m\to\infty} \left[ \frac{m - 1 + \gamma}{m} \right] r_m, \;\;\; r_m=\frac{a_{m-1}}{a_m},
\label{U_c}
\end{equation}
given a limited number $m_*$ of the series coefficients. We find that $m_*$ currently accessible by state-of-the-art DiagMC algorithms is already sufficient for a controlled determination of $U_c$ of the N\'eel transition with acceptable error bars. 

\begin{figure}
\includegraphics[width=1\columnwidth]{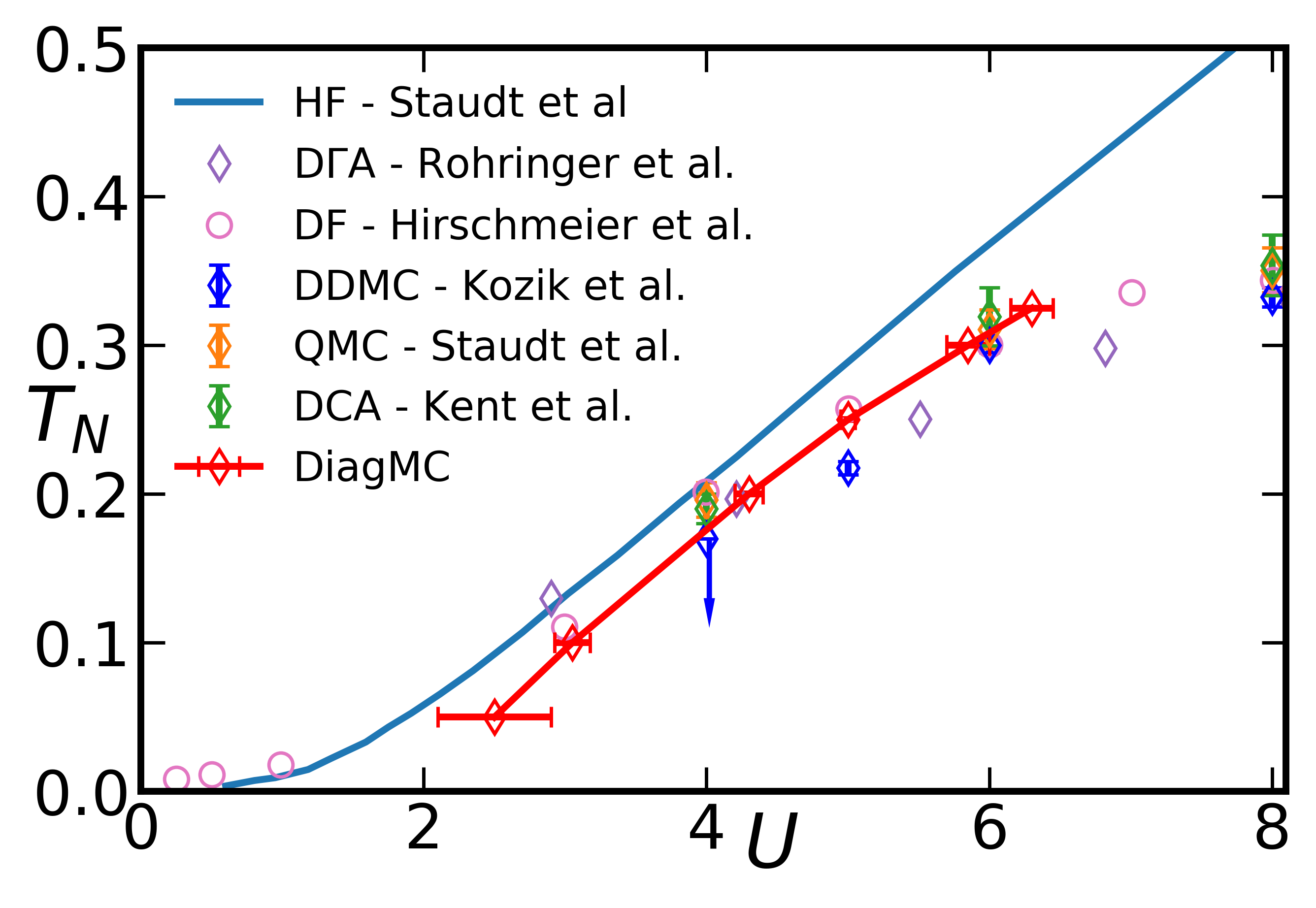}
\caption{\label{benchmark} {A comparison of our DiagMC results for the N\'eel temperature $T_N(U)$ (in $t=1$ units) at half-filling with those of state-of-the-art controlled and approximate methods (see text).}}
\end{figure}

A natural choice of the observable for this problem is the spin structure factor
\begin{equation}
S(\mathbf{q})= \frac{1}{N} \sum_{\mathbf{x}_i, \mathbf{x}_j} e^{i\mathbf{q} \cdot (\mathbf{x}_i - \mathbf{x}_j)} \langle S^{z}_{i}S^{z}_{j} \rangle, 
\label{S_q}
\end{equation}
where $S^{z}_{i}=(\hat{n}_{i, \uparrow}-\hat{n}_{i, \downarrow})/2$ is the spin projection operator at the lattice site $\mathbf{x}_i$ and $N \to \infty$ is the total number of lattice sites. On the one hand, $S(\mathbf{q})$ can be measured \textit{in situ} with UCA in optical lattices~\cite{Miyake_2011,Hart_2015, Mazurenko2017} and thus is useful for benchmarking and calibrating experiments on the Hubbard model. On the other hand, $S(\mathbf{q})$ at $\mathbf{q}=\boldsymbol{\pi} \equiv (\pi, \pi, \pi)$ diverges at the AFM N\'eel transition and shares its critical index with the divergent static AFM susceptibility $\chi(\omega=0,\boldsymbol{\pi})$ since $S(\mathbf{q}) = \sum_{\omega}{\chi(\omega,\mathbf{q})}$, with $\chi(\omega=0,\mathbf{q})$ being the dominant part. Specifically, $S(\boldsymbol{\pi}) \propto (T-T_N)^{-\gamma}$, where $\gamma = 1.396(9)$ is the critical index of the $3d$ Heisenberg universality class~\cite{heisenbergExponents}. Since $T_N(U, \mu)$ is a smooth function, for $U \to U_c$ we have $S(\boldsymbol{\pi}) \propto (U_c-U)^{-\gamma}$, and thus $U_c$ can be determined by Eq.~\ref{U_c} from the series coefficients $a_m$ for $S(\boldsymbol{q}) = \sum_m a_m U^m$~\footnote{Our algorithm actually computes $a_m=a_m(\mathbf{q}, T, \tilde{\mu})$ for a fixed \textit{shifted} chemical potential $\tilde{\mu}$ such that $\mu=\tilde{\mu} + \nu U$ with some constant $\nu$~\cite{Rossi2016det, Simkovic2019}, but $T_N(U, \tilde{\mu})$ is equally smooth.}.

\begin{figure}
\includegraphics[width=1\columnwidth]{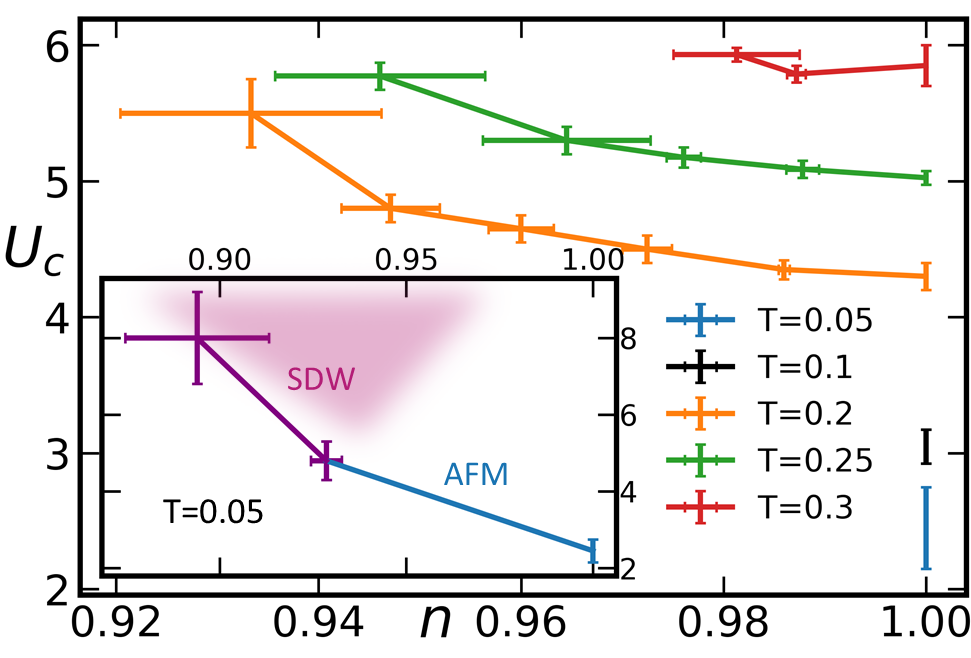}
\caption{\label{doped} The critical value of interaction $U_c$ of the N\'eel transition at fixed temperatures $T$ as a function density $n$ (in $t=1$ units). The inset shows the data at $T=0.05$, where the purple points indicate the transition to an incommensurate SDW state. }
\end{figure}

We employ the CDet algorithm~\cite{Rossi2016det}, adapted to the magnetic structure factor $S(\boldsymbol{q})$~\cite{Kim2020spin_charge}, to compute $a_m$ up a maximum expansion order $m_* \sim 8$. Figure \ref{adjusted_ratio}(a) illustrates the application of Eq.~\ref{U_c} by showing the dependence of the ratios $r_m=\frac{a_{m-1}}{a_{m}}$ on the diagram order $m$ for $T=0.25t$, $n=1.0$, and $\mathbf{q}=\boldsymbol{\pi}$. The comparison of $r_m$ with its counterpart $\frac{m U_c}{m - 1 + \gamma}$ for the function $(U-U_c)^{-\gamma}$ suggests that we have reached the asymptotic regime with $U_c \approx 5.0$ beyond $m \sim 5$. This justifies the use of the adjusted ratios $\tilde{r}_m=(m-1+\gamma) r_m/m$ in Eq.~\ref{U_c} to identify $U_c$ as the location of the plateau of $\tilde{r}_m$ at large $m$. Figure \ref{adjusted_ratio}(b) demonstrates this protocol, where the error bars are judged by the overlap between the error bars of $\tilde{r}_m$ for $m \geq 6$. In Fig.~\ref{adjusted_ratio}(c), we show the behavior of the adjusted ratio $\tilde{r}_m$ at higher temperatures ($T=0.5$ at $n=1$ in this example), where there is no N\'eel order at any $U$: the ratios do not converge to a plateau since the singularities closest to the origin in the complex plane of $U$ are not on the real axis. For some parameters at larger doping we also observe a singularity at or near $U<0$, likely due to a transition to a superconducting state in the attractive Hubbard model~\cite{Rossi2016det}. This singularity, however, can straightforwardly be eliminated by a simple conformal map of the series $\sum_m a_m U^m$~\cite{Simkovic2019}, or, equivalently, by a homotopy of the effective action~\cite{Kim2021homotopy}, so that the resulting coefficients $a_m$ display only the physical singularity at $U_c$ in Eq.~\ref{U_c}.

We first use the protocol of Fig.~\ref{adjusted_ratio}(b) to evaluate $U_c(T)$ of the N\'eel transition at half-filling.  Figure~\ref{benchmark} benchmarks our data against the established controlled results of QMC~\cite{Staudt2000}, DCA~\cite{Kent2005}, and DDMC~\cite{Kozik2013Neel}, as well as the Hartree-Fock approximation~\cite{Staudt2000}, and those of advanced approximate methods: D$\Gamma$A~\cite{Rohringer2011} and dual fermions (DF)~\cite{Hirschmeier2015}. 
The curve $T_N(U)$ is known~\cite{Staudt2000, Kent2005, Kozik2013Neel} to be non-monotonic, with a broad maximum around $U \sim 6-8t$, where $T_N \approx 0.33t$, due to the different mechanisms of the N\'eel transition in the $U/t \ll 1$ (Slater~\cite{Rev:1951ib}) and $U/t \gg 1$ (Mott-Heisenberg) limits. Thus the $U_c(T)$ dependence is non-single-valued and our technique, when applied to the Hamiltonian~(\ref{Hubbard}), can only capture the Slater-like branch closest to the $U=0$ starting point of the diagrammatic expansion. It should be possible to detect the other branch with our technique applied to an equivalent Hamiltonian or effective action transformed by a suitable homotopy~\cite{Kim2021homotopy} that modifies the trajectory of the model on its way from $U=0$ to $U=U_c$. Nonetheless, in the present form our method is efficient at locating the transition in the most correlated regime around the broad maximum (around $T_N = 0.325t$ at $U=6.30(15)t$ in our calculation), where our results agree with those of the finite-size scaling approaches with similar error bars, down to temperatures $T<0.2t$ where no other controlled data are available. At $T \gtrsim 0.33t$, the AFM structure factor does not display a transition at any $U$, also in consistency with the previous studies. Curiously, the DF data~\cite{Hirschmeier2015} follow the $T_N(U)$ curve within error bars for all temperatures down to $T=0.05$, where our accuracy deteriorates (generally at all densities) due to larger error bars of $a_m$ and the plateau in Eq.~\ref{U_c} appearing at larger $m$.

\onecolumngrid 
 
\begin{figure}[h!]
\centering
\includegraphics[width=0.95\textwidth]{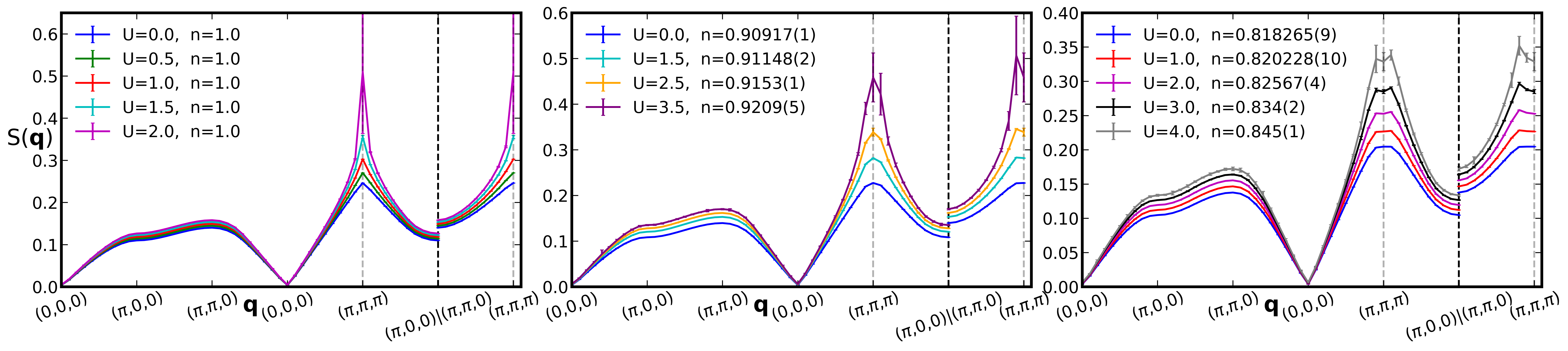}
\caption{\label{brillouin_zone} {Evaluation of $S(\mathbf{q})$ at $T=0.05t$ at several densities $n$ and coupling $U$ ($t=1$) on approach to the N\'eel transition in the inset of Fig.\ref{doped}. The peak of $S(\mathbf{q})$ moves from $\boldsymbol{\pi}$ to $\mathbf{q}_\mathrm{SDW}$ with doping, indicating a shift to dominant incommensurate SDW correlations.}}
\end{figure}
\twocolumngrid

The absence of the sign problem in our approach gives it a significant advantage in the doped case, allowing a controlled identification of the critical point with no reduction in the precision. The AFM phase diagram has previously been studied by the second-order perturbation theory (with an additional local approximation)~\cite{PhysRevB.55.942} and more recently using the state-of-the-art D$\Gamma$A approximation~\cite{Schaefer2017}. Both studies demonstrate a gradual reduction of $T_N$ with doping at fixed $U$ and that at sufficiently large doping beyond $5-15\%$ (depending on $U$) the N\'eel transition is to an incommensurate SDW state with $S(\mathbf{q})$ diverging for $\mathbf{q}=\mathbf{q}_\mathrm{SDW} \equiv (\pi, \pi, q_z <\pi)$ instead of $\boldsymbol{\pi}$. The $T_N(n)$ line was shown to reach $0$ at $10-20\%$ doping at a magnetic quantum critical point (QCP). Ref.~\cite{Schaefer2017} further reveals a peculiar scenario of criticality of the N\'eel transition in the doped system, essentially driven by features of the Fermi surface (FS): As the FS is deformed by doping and the ordering with the AFM nesting vector $\boldsymbol{\pi}$ transitions to that with the incommensurate $\mathbf{q}_\mathrm{SDW}$ connecting the Kohn points, the critical indices of the paramagnetic-to-SDW transition remain those of the $3d$ Heisenberg universality class. However, as $T_N$ continues to drop with further doping and the system enters the quantum critical regime, the critical indices cross over to those of the QCP controlled by the Kohn anomaly, with $\gamma = 0.5$ in particular. 

Fig.~\ref{doped} presents our results for the dependence of $U_c$ on density $n$ (for several temperatures). The values of $n$ are obtained by a controlled extrapolation of DiagMC data to infinite order at $U_c$ following Ref.~\cite{Simkovic2019} and using the Dlog-Pad\'{e}~\cite{baker1961Dlog} method, enabled by the convergence of the perturbative expansion for $n$ at $U_c$. As expected from Refs.~\cite{PhysRevB.55.942, Schaefer2017}, $U_c$ rises (at fixed $T$) with doping, which can be attributed to a stronger interaction being required in the doped system to suppress the double occupancy in the Slater AFM mechanism. This qualitative behavior is observed, e.g., in (La$_{1-x}$Ba$_x$)$_2$CuO$_4$~\cite{realNeelTemp}, where doping decreases the N\'eel temperature $T_N$. The $U_c(n)$ curve for a fixed $T$ ends naturally at the density $n_*$ where the function $T_N(U)|_{n=n_*}$ reaches its maximum, so that there is no N\'eel transition for $n<n_*$ at any $U$. Since this maximum is typically broad (cf. Fig.~\ref{benchmark}), the error bars of $U_c$ generally grow on approach to $n_*$. At $T=0.05t$, and $n \lesssim 0.93$, we observe that $S(\mathbf{q}_\mathrm{SDW})$ diverges before, i.e. at a smaller $U$ value, than $S(\boldsymbol{\pi})$ beyond the error bars, indicating that the N\'eel transition is to an incommensurate SDW. Large error bars at low $T$ prevent reliable estimates of the critical index $\gamma$, which appears roughly consistent with the $3d$ Heisenberg criticality in this regime~\cite{Schaefer2017}. The QCP regime at $n \sim 0.8$ and $T_N< 0.05t$, predicted by D$\Gamma$A~\cite{Schaefer2017}, remains challenging and deserves a separate study.

Fig.~\ref{brillouin_zone} illustrates the development of magnetic correlations on approach to the transition line at $T=0.05t$ (inset of Fig.~\ref{doped}): $S(\boldsymbol{q})$, obtained by the same controlled extrapolation as for $n$~\cite{Simkovic2019, baker1961Dlog}, is shown along the high-symmetry path through the Brillouin zone for several $n$ and increasing values of $U$. As expected, a sharp peak at the AFM wavevector $\boldsymbol{\pi}$ develops at half-filling. This peak is broadened by small doping $\sim 10\%$, at which a second peak at $\mathbf{q}_\mathrm{SDW}$ starts developing with increasing $U$. The SDW peak surpasses the AFM one already for $U>2t$ at $n \approx 0.91$. At larger doping $\sim 10-15\%$, $S(\boldsymbol{q})$ still starts with a single AFM peak at $U=0$, albeit further broadened to cover $\mathbf{q}_\mathrm{SDW}$, but increasing $U$ leads to the build up of SDW correlations directly, without a noticeable AFM-to-SDW crossover. These results resemble the crossover to incommensurate magnetic correlations observed in the $2d$ Hubbard model \cite{vsimkovic2021weak} and provide controlled benchmarks for UCA experiments showing a regime where profitable experiments observing changes in the dominant spin correlations could be carried out. 

In summary, the breakdown of diagrammatic expansions is a practical tool of detecting and characterizing second-order phase transitions at temperatures where the series coefficients $a_m$ at high ($m_* \gtrsim 8$) orders can be computed with sufficient accuracy. Its formulation directly in the TDL circumvents the fermion sign problem of finite-size methods and enables access to new regimes. It is complementary to and substantially simpler in practice than the technique of detecting the transition through DiagMC evaluation of the eigenvalues of relevant vertex functions~\cite{Gukelberge2014nematic, deng2015emergent}, which, however, seems unavoidable whenever the transition happens at very low temperatures (relative to the Fermi energy)~\cite{deng2015emergent, simkovic2021superfluid, implicit_renormalisation}. The singularity analysis can in principle be used for detecting first-order transitions as well, but is likely less practical than the direct evaluation of the free energy for both phases~\cite{spada2021highorder} since the metastability below a first-order transition manifests itself in an essential singularity, typically revealed at high orders. Perturbative expansions in $U$ about the free Fermi gas are generally thought to be applicable at weak-to-moderate coupling; our results evidence that, when powered by DiagMC, they are a precise tool at least until the nearest phase transition.

\begin{acknowledgments}
This work was partially supported by EPSRC through the grant EP/P003052/1 and by the Simons Foundation as a part of the Simons Collaboration on the Many Electron Problem. We are grateful to the UK Materials and Molecular Modelling Hub for computational resources, which is partially funded by EPSRC (EP/P020194/1).
\end{acknowledgments}



\bibliography{refs}

\end{document}